\documentclass{llncs}
\usepackage{amsmath, amssymb}
\usepackage{graphicx}
\usepackage{multirow}
\usepackage[latin1]{inputenc}

\newcommand{\remove}[1]{}

\newenvironment{changemargin}[2]{%
  \begin{list}{}{%
    \setlength{\topsep}{0pt}%
    \setlength{\leftmargin}{#1}%
    \setlength{\rightmargin}{#2}%
    \setlength{\listparindent}{\parindent}%
    \setlength{\itemindent}{\parindent}%
    \setlength{\parsep}{\parskip}%
  }%
  \item[]}{\end{list}}

\newcommand{\newsel}{{\sc super-selector}}

\def\x{\textbf{x}}
\def\y{\textbf{y}}
\def\z{\textbf{z}}
\newcommand{\X}{\tilde{X}}
\def\v{{\bf v}}

\newcommand{\commento}[1]{\marginpar{\tiny \flushleft{#1}}}

\newtheorem{fact}{Fact}
\begin{document}

\title{Superselectors: Efficient
Constructions and Applications}

\author{Ferdinando Cicalese \and Ugo Vaccaro}

\institute{Department of Computer Science and Applications ``R.M. Capocelli''\\
University of Salerno, via Ponte don Melillo, 84084 Fisciano, Italy}

\maketitle

\begin{abstract}
We introduce a new combinatorial  structure: \textit{superselectors}.
We show that superselectors  subsume several 
important combinatorial structures used in the past few years  to solve problems 
in group testing, compressed sensing, multi-channel conflict resolution and  
data security. We prove  close upper and lower bounds on the
size of superselectors and we provide efficient algorithms
for their constructions. 
Albeit  our bounds are very general, when they are instantiated on the combinatorial 
structures that are particular cases of superselectors
(e.g., $(p,k,n)$-selectors \cite{DGV}, $(d,\ell)$-list-disjunct matrices \cite{soda10}, $MUT_k(r)$-families \cite{LR}, 
$FUT(k, \alpha)$-families \cite{Alon-Hod}, etc.) they 
match the best known   bounds in terms of size of
the structures (the relevant parameter in the applications).
For appropriate values of  parameters, our results also provide the 
first efficient deterministic
algorithms for the construction of such structures.
\end{abstract}

\section{Introduction}

It is often the case where understanding and solving a problem
means discovering the \textit{combinatorics} at the heart of
the problem. Equally time and again it happens that the crucial step towards 
the economical solution of  problems arising in different areas
hinges on the efficient construction of a \textit{same} combinatorial object. 
An interesting example is that of superimposed codes \cite{kts}
(also known as cover-free familes \cite{eff}, strongly selective families \cite{cms}, disjunct
matrices \cite{du}, ...).
Superimposed codes represent the main  tool for the efficient solution of several
problems arising in compressed sensing \cite{CM}, cryptography and data security \cite{black},
computational biology \cite{balding},
 multi-access communication \cite{Wolf},  database theory
\cite{kts}, pattern matching \cite{ind,Porat_ESA,Porat_FOCS},  distributed
colouring \cite{linial}, and  circuit complexity \cite{char}, among the others.
Due to their importance, a lot of efforts has been devoted to the design of
fast algorithms for the construction of superimposed codes of short length. In this
line of research a main result is  the paper by Porat and Rotschild \cite{PR}
who presented a very efficient polynomial time algorithm for that purpose.
More recently,  Indyk {\em et al.} \cite{soda10}  showed that optimal nonadaptive group testing procedure  
(i.e, superimposed codes) can be efficiently constructed and decoded. 

In the past few years it has also become apparent that combinatorial structures 
strictly related to superimposed codes lie at the heart of an even more vast 
series of problems. As quick examples, the selectors introduced in \cite{CGR}
were instrumental to obtain fast broadcasting algorithms in radio networks, the 
$(p,k,n)$-selectors of \cite{DGV} were the basic tool for the first
two-stage group testing algorithm with an information theoretic optimal
number of tests, the $(d,\ell)$- disjunct matrices of \cite{soda10} were a crucial building block
for the efficiently decodable non-adaptive group testing procedures mentioned above.

It is the purpose of this paper to introduce \textit{superselectors}, a new combinatorial object 
that encompasses and unifies all of the combinatorial structures mentioned above (and  more).
We provide efficient methods for their  constructions and  apply their  properties 
to the solutions of  old and new problems for which constructive 
solutions have not been shown so far.
In particular, superselectors extend at the same time superimposed codes and 
several different  generalizations of theirs proposed in the literature.

When appropriately instantiated, our superselectors asymptotically match 
the best known constructions of 
$(p,k,n)$-selectors \cite{DGV}, $(d, \ell)$-list-disjunct matrices \cite{soda10}, 
monotone encodings and $(k, \alpha)$-FUT families \cite{Moran,Alon-Hod}, 
$MUT_k(r)$-families  for 
multiaccess channel \cite{LR,Alon-Asodi}. In some cases, e.g., for $(p,k,n)$-selectors and $(d, \ell)$-list-disjunct
matrices,   
we also improve on the multiplicative constant in the O  notation.
We show that optimal size superselectors (and hence all the above structures) can 
be easily constructed in time polynomial in $n,$  the main dimension of the structure, though exponential in the second 
parameter $p.$ This might be satisfying in those applications, e.g., computational biology, where 
$p \ll n$.  A major  open question is whether it is possible to deterministically obtain optimal size superselectors  (or even  selectors)
in time subexponential in $p.$
However,  in   
cases when   
$p$ is constant 
 we note that our results provide the first known polynomial construction of 
\emph{optimal size} $(p,k,n)$-selectors (and related structures). 

It should be also noticed that selectors, and similar combinatorial structures, generally have  to
be computed \emph{only once}, since they can be successively used in different
contexts   without the need to recompute them from scratch.
Therefore, it seems to make sense (and in absence of  better alternatives) to have
onerous  algorithms that output structures of 
optimal size, (the crucial parameter that will affect the complexity of algorithms 
that uses selectors and the like structures in different 
scenarios) than more efficient construction algorithms that 
produce structures of suboptimal size.
This brings us to another question.
Most of the structures  mentioned above, and subsumed by our
 superselectors,
  can  also be obtained via 
expander graphs, or equivalently, randomness extractors. However, to  the best of our knowledge, the best known explicit
expander-based constructions give only suboptimal (w.r.t. to the size) selector-like structures.
Table \ref{table:comparisons} summarizes how our 
results compare to the state of the art. 
The bounds are reported as they were given in the original papers, thus producing a slight 
 level of difformity. 
However, if with this choice we might be requiring  the reader to put a little bit of 
effort in the comparisons, we are not risking mistranslations 
of the bounds from one notation into another.
The main aim of the data in the table is  to show that 
the generalization provided by
the superselectors in no case implies a loss in terms of optimality of the structure size.
In addition, the number of applications of superselectors 
we shall present in Section 3 seems to suggest that 
they represent a basic structure, likely to 
be useful in many 
contexts.

\begin{table}[ht]
\begin{changemargin}{-2.3cm}{+2.3cm}
{\small
\begin{center}
\begin{tabular}{|l||l|l|l|} 
 \hline
 Structure
  &  Lower Bounds on the Size
       &    Our Upper Bounds on the Size , \!\!\!
        &  Old Upper Bounds on the Size,  \\ 
   & & construction time & construction time \\
        \hline \hline
\multirow{2}{*}{$(p,k,n)$-sel. }& 
$\Omega\left(\frac{p^2}{p-k+1} \frac{\log(n/p)}{\log\frac{p}{p-k+1}+O(1)}\right)$ \cite{Chlebus,DGV} \!\!\!& 
$\frac{2p^2}{p-k+1} \log\frac{n}{p} \; (1+o(1))$ & 
$\frac{p^2}{p-k+1} \, polylog(n),$  time:  $poly(n)$ \cite{Chlebus} \\
& &time: $O(n^{p+1}\log n)$ & $\frac{e p^2}{p-k+1} \log \frac{n}{p} + \frac{ep(2p-1)}{p-k+1},$ time: $EXP(n)$ \cite{DGV}
\\ \hline
\multirow{3}{*}{$(d, \ell)$-list}&
$d \log \left(\frac{n}{e(d + \ell -1)}\right)$  if $d<2\ell$ &
$O\left(2d \log \frac{n}{2d} \right),$ $d < \ell$& 
$O\!\!\left(\frac{2(d+\ell) \log n + \log {{d+\ell}\choose {d}}}{\frac{\ell}{d+\ell} \left(\frac{d}{d+\ell}\right)^{d/\ell}}\!\right),$  \\ 
& $\frac{d^2}{4\ell \log\frac{e d^2}{ 4 \ell}} \log \frac{n-2\ell-\frac{d}{2}}{e d^2}$ if $d \geq 2\ell$  \cite{DGV} & 
 $O\left(\frac{(d+\ell)^2}{\ell} \log \frac{n}{d+\ell}\right),$  $d \geq \ell$
&     
time: $poly\left({{d+\ell} \choose {d}}, n^{d+\ell}, 2^{d+\ell}, \frac{ \left(\frac{d}{d+\ell}\right)^{-\frac{d}{\ell}}}{\frac{\ell}{d+\ell}}\right)$  
\cite{soda10}
\\ 
& & time: $O(n^{d + \min\{\ell, d\} + 1}\log n)$& \\ \hline
%
%
$MUT_k(p)$ & $\Omega\left(\max\{k^2, p\} \log \frac{n}{p}\right)$ \cite{LR} & 
 $O((p+k^2) \log \frac{n}{p})$ &
  $O((p+k^2)\log \frac{n}{p}),$ non-constructive \cite{Alon-Asodi} \\ 
  & & time: $O(n^{p+1}\log n)$& \\
  \hline
$(p,\alpha)$-FUT   &  
$\frac{p}{(1-\alpha)\log \frac{p}{1-\alpha}}  \log n$ \cite{Alon-Hod}&     
$\frac{p}{(1-\alpha)}\log \frac{n}{p}$ &
$\frac{p}{(1-\alpha)}\log \frac{n}{p}$,  non-constructive   \cite{Alon-Hod} \\ 
  & & time: $O(n^{p+1}\log n)$& 
  \\ \hline
\multirow{3}{*}{$p$-cover free}& 
$\left(\frac{p^2}{2 \log p} \log\frac{n}{p}\right)\!(1\!+\!o(1))\!$ \cite{DR} \!\!\!& 
$\frac{e p^2}{\log e} \log \frac{n}{p}\! (1\!+\!o(1))\!$ \!\!& 
$p(p+1) \log e  \log \frac{n}{p},$ non-constructive  \cite{Dyachkov,Cheng}
\\ 
 & & time: $O(n^{p+1}\log n)$ &\\ 
 & & & $\Theta(p^2 \log n),$ time: $\Theta(p n \log n)$ \cite{PR}. \\
 \hline \hline
$(p,{\bf v}, n)$-sel.   &  
$\max_{j} \frac{j^2}{j-v_j+1} \frac{\log(n/j)}{\log \frac{j}{j-v_j+1} + O(1)}$&   
 \multicolumn{2}{l|}{$\max_{j=1,...,p} \{
\min \{ \frac{j^2}{\log_2 e},  \frac{3pej}{j-v_j+1}\}\log\frac{n}{j}\}$ }
\\ \hline
\end{tabular}
\end{center}

}
\end{changemargin}
\caption{\small {\bf Bounds attained via  \newsel{\sc s} against best known bounds.} \label{table:comparisons}}

\end{table}
\normalsize
%
%
%
\vspace*{-0.8truecm}
\section{The  $(p,\textbf{v},n)$-\newsel} \label{sec:strong_selectors}

\vspace*{-.1truecm}
\noindent 

Given two vectors ${\bf x}, {\bf y} \in \{0,1\}^n,$ we denote with 
${\bf x} \oplus {\bf y} $ the Boolean sum of ${\bf x}$ and ${\bf y},$ i.e.,  their componentwise OR.
Given an $m \times n$ binary matrix $M$ and an $n$-bit  vector ${\bf x},$
we denote by  $M \odot {\bf x}$ the $m$-bit vector obtained
by performing the Boolean sum of the columns of $M$ corresponding to the  
positions of the  $1$'s in $\bf x.$ That is, if $\bf x$ has a $1$ in positions, say $3,7,11, \dots,$
then $M \odot {\bf x}$ is obtained by performing the $\oplus$ of the $3$rd, $7$th, $11$th, \dots, 
column of $M.$
Given a set $S \subseteq [n],$ we use $M(S)$ to denote the submatrix induced by the columns with index in $S.$
Also we use ${\bf a}_S$ to indicate the Boolean sum of the columns of $M(S).$
Given two  $n$-bit vector ${\bf x}, {\bf y}$ we say that $\bf x$ is {\em covered} by $\bf y$ if 
$x_i \leq y_i,$ for each $i = 1, \dots, n.$  Note that if $\bf x$ is not covered by $\bf y$ then 
it means that $\bf x$ has a $1$ in a position in which $\bf y$ has a $0.$

\remove{Following \cite{DGV}, in this paper, we will prefer to use the following definition of a $(p,k,n)$-selectors, which is 
equivalent to the one given in the introduction in terms of families of sets}
We first recall the definition of $(p, k, n)$-selector, as given in \cite{DGV}.
A $(p, k, n)$-selector is an $m \times n$ binary matrix such that for any subset $S$ of $p\leq n$ columns, 
the submatrix $M(S)$ induced by $S$ 
contains at least $k\leq p$ rows
of the identity matrix $I_p$.
The parameter  $m$ is the \textit{size} of the selector.

\begin{definition}
Fix integers $n, p,$ with $p \leq n$ and an integer vector, $\v = (v_1, \dots, v_p),$ such that 
$v_i \leq i,$ for each $i =1, \dots, p.$ 
We say that an $m \times n$  binary  matrix $M$ is a  $(p , \v, n)$-\newsel\  if
$M$ is a $(i, v_i, n)$-selector for each $i = 1, \dots, p.$
We call $m$ the size of the \newsel.
\end{definition}

Our main result on \newsel{\sc s} is summarized in the following theorem, whose proof will be given in 
Sections \ref{sec:construction}. 

\begin{theorem} \label{theo:main}
A $(p,\v,n)$-\newsel\ of size
$$m = O(\max_{j=1, \dots, p} k_j \log (n/j)), \qquad \mbox{where } 
k_j =  \min\left\{\frac{3pej}{(j-v_j + 1)}, \frac{e j^2}{\log_2 e} \right\}$$
can be constructed in time polynomial in $n$ and exponential in $k.$

\end{theorem}

The  ``identification'' capability of a \newsel\ are as follows.

\begin{lemma} \label{lemma:supsel_prop}
Let $M$ be a $(p, \v, n)$-\newsel, $\v=(v_1, \ldots , v_p)$. 
Let  $S$ be any set of $x < v_p$ columns of $M.$ Let ${\bf a}_S$ denote the 
Boolean sum of the columns in $S.$ Then, from ${\bf a}_S$ it is possible to identify at least $v_{x+y}$ of the columns in $S,$
where $y$ is the number of columns of $M$ which are not in $S$ but are covered by ${\bf a}_S.$ Moreover, 
$y < \min\{j \mid x < v_j\} - x.$
\end{lemma}
\begin{proof}
Let $T = \{ {\bf b} \mid {\bf b} \not \in S~{\rm and}~{\bf b} \oplus {\bf a}_S = {\bf a}_S\},$ i.e., $T$ is the set of columns not in $S$ but 
covered by ${\bf a}_S.$ Then, $y = |T|.$ We first prove the last statement.

\noindent
{\em Claim.} $y < \min\{j \mid v_j > x\} - x.$ Let $j^*$ be a value of $j$ achieving the minimum.
The claim is a consequence of  $M$ being a $(j^*, v_{j^*}, n)$-selector. To see this, assume, by contradiction, that $|T| \geq j^* -x.$ 
Let $T'  \subseteq T$ and $|T' \cup S|  = j^*.$ Then, 
there are at least $v_{j^*} > |S|$ columns in $T' \cup S$ with a $1$ in a row where all the other columns have a $0.$ Thus,  
there is at least  one column of $T'$ which has a $1$ where all the column of $S$ have a $0.$ This contradicts the fact that all the 
columns of $T$ (and hence of $T'$) are covered by ${\bf a}_S.$


Since $x+y < j^* \leq p,$ and $M$ is an $(x+y, v_{x+y}, n)$-selector, among the columns of $S \cup T$ there are 
at least $v_{x+y}$ which have a $1$ where all the others have a $0.$ Let $W$ be such set of columns.  
By an argument analogous to the one used in the claim  we have that $W \subseteq S$ and we can identify them.\qed
\end{proof}

\begin{remark}
Notice that if  $v_i > v_{i-1}, $ for each $i=2, \dots, p,$ then we have a situation that, at a first look, 
might appear surprising: the larger is
the number of spurious elements, i.e., columns not in $S$ but covered by ${\bf a}_S,$ the 
more information we get on $S,$ i.e., the 
more are the columns of $S$ that are identified.
\end{remark}

\begin{remark} \label{remark:additive}
The same argument used in the proof above shows that Lemma \ref{lemma:supsel_prop} also holds when ${\bf a}_S$ is the 
component-wise arithmetic sum of the columns in $S.$
\end{remark}

\section{Applications of the \newsel{\sc s}}

\noindent
{\bf Approximate Group Testing.} 
In classical non-adaptive group testing \cite{du},  we want to identify a subset $P \subseteq [n],$ with $|P| \leq p,$
by using the minimum possible set of tests $T_1, \dots, T_m,$ where for each $i=1, \dots, m,$ 
we have $T_i \subseteq [n].$ The outcome of test $T_i$ is a bit which is $1$  iff  $T_i \cap P \neq  \emptyset.$ 
If we require that the whole $P$ is identified exactly, 
and non-adaptively, then it is known that $\Omega(\frac{p^2}{\log p} \log \frac{n}{p})$
tests are necessary \cite{du}. 

Cheraghchi \cite{Cheraghchi}, in the context of error-resilient group testing,  Gilbert {\em et al.} \cite{Gilbert}, in the context of 
sparse signal recovery,  and Alon and Hod \cite{Alon-Hod} considered 
the case when one is interested  in identifying some approximate version of $P.$ It turns out \cite{Cheraghchi} 
that at least $p \log \frac{n}{p} - p - e_0 - O(e_1 \log\frac{n-p-e_0}{e_1})$ tests are necessary 
if one allows the identification algorithm to report a set $P',$ such that
$|P'\setminus P| \leq e_0$ and $|P\setminus P'| \leq e_1.$ 
In other words, the algorithm can report up to $e_0$ false positives and up  to $e_1$ false negatives.

Let $M$ be an appropriate  $(p+e_0, \v, n)$-\newsel, with the components of vector 
$\v$ defined by $v_i = i- \min\{e_0, e_1\}+1.$ We can use $M$ to attain approximate identification in the 
above sense.
Proceeding in  a standard way, map $[n]$ to the indices of the columns of the super-selector and interpret the  
rows of the super-selector as the indicator vectors of the tests. 
Now the vector of the outcomes of the tests is the Boolean sum ${\bf a}_P$ of those
columns whose index is in $P.$  Let $P'$ be the set of the indices of the columns covered by ${\bf a}_P.$ We have $P \subseteq P'$ and 
by Lemma \ref{lemma:supsel_prop} also $|P'| \leq |P| + e_0.$ Moreover, from Lemma \ref{lemma:supsel_prop} we also know that
a set of positives $P'' \subseteq P$ can be  exactly identified, with   $|P''| \geq |P| - e_1.$  
Therefore, any set $P^*$ with $P'' \subseteq P^* \subseteq P'$ satisfies the bounds on the false positives and false negatives.

Note that, for the interesting case of $e_0, e_1 = \Theta(p)$, the above group testing strategy is 
 best possible since it uses
 $O(p \log \frac{n}{p})$ tests which  matches the lower bound of \cite{Cheraghchi}.
 Cheraghchi \cite{Cheraghchi} considers the case when some tests migh be erroneous and only
 focuses on the case of zero false negatives. 
 Alon and Hod \cite{Alon-Hod} consider the case of zero false positives and obtain  
 $O(p\log (n/p))$ tests procedures, which are in fact optimal for this case.
 Gilbert {\em et al.} \cite{Gilbert} allow both false positives and false negatives
 but their procedures uses $O(p\log ^2n)$ tests.
 Moreover, our implementation  guarantees the exact identification 
 of at least $p'-\min\{e_0,e_1\}+1$ positives, where $p'\leq p$ is
 the actual number of positive elements.

\smallskip

\noindent
{\bf Additive Group Testing.} 
We now consider exact   group testing with {\em additive} tests. In this variant,  
 the outcome of testing a subset $T_i$ 
is the number of positives contained in $T_i,$ i.e., the integer $|T_i \cap P|.$ 

It is known that $\Omega(\frac{p}{\log p} \log \frac{n}{p})$
tests are necessary if we want to exactly identify $P$ using additive tests (see, e.g., \cite{GK} and references therein). 

Proceeding analogously to the case of Approximate Group Testing, we can reformulate 
the additive group testing problem 
as follows: given positive integers $n$ and $p < n,$ minimize the number $m$ of rows of an 
$m \times n$ $0$-$1$ matrix  $M$ such that any set $P$ of up to $p$ columns of $M$ can be  
identified  from their  
sum\footnote{Here  sum  is  meant in the arithmetic way, i.e., 
${\bf z} = {\bf x} + {\bf y}$ iff $z_i = x_i + y_i,$ for each $i.$}${\bf a}_P.$ \enlargethispage{0.3cm} 

Let $M$ be an appropriate  $(2p, \v, n)$-\newsel, with the components of vector 
$\v$ defined by $v_i = i,$ for $i=1, \dots,  \sqrt{p}$ and $v_i = \lceil \frac{i}{2} \rceil + 1,$ for $\sqrt{p} < i \leq 2p.$
We show that $M$ provides a non-adaptive strategy for additive group testing with  $O(p \log (n/p))$ tests. 

If $|P| < \sqrt{p},$ using the fact that $v_{|P| + 1} = |P|+1,$ Lemma \ref{lemma:supsel_prop} and Remark \ref{remark:additive}
imply that from ${\bf a}_S$ we can identify the whole set $P.$

If, otherwise, $|P| \geq \sqrt{p},$ by using the fact that $v_{2|P|} > |P|,$ 
by Lemma \ref{lemma:supsel_prop} and Remark \ref{remark:additive}, 
from ${\bf a}_P$ we  can uniquely identify a subset $R$ of $P,$ such that $|R| \geq p/2$ and 
confine the elements of $P_1 = P \setminus R$ into a set $S_1$ such that 
$|S_1| \leq p.$ In particular $S_1 \cup R$ is the set of all columns of $M$ which are component-wise 
not larger than ${\bf a}_P.$

Now,  let ${\bf a}_{P_1} = {\bf a}_P - \sum_{i \in R} {\bf c}_i,$ where ${\bf c}_i$ 
denotes the $i$th column of $M$ and 
the additions and subtractions among vectors are meant component-wise. Clearly, 
${\bf a}_{P_1}$ is the sum of $P_1,$ i.e., the columns that are still to be identified. Note also that 
${\bf a}_{P_1}$ can be computed from ${\bf a}_P$ and the set $R$ of identified columns \emph{without} 
any additional test. 

We have now  a smaller instance of the same problem from which we started, namely 
identifying the columns of $P_1,$ among the ones in $M(S_1 \setminus R),$
from their sum ${\bf a}_{P_1}.$ Also notice that Lemma \ref{lemma:supsel_prop} still applies to the 
columns of  $M(S_1 \setminus R).$ 
Therefore,  repeatedly using the above argument we can eventually identify the whole
set $P.$ Again, no additional tests are required since we 
reinterpret, so to speak, the tests outcomes in light of
new acquired knowledge.

Finally, by Theorem \ref{theo:main} a \newsel\ $M$ of size $O(p \log \frac{n}{p})$ can be constructed 
in time $O(n^p),$ which gives the desired result.
We hasten to remark that in  \cite{GK}  Grebinsky and Kucherov prove the existence of matrices $M$ with an
optimal $O(\frac{p}{\log p} \log \frac{n}{p})$ number of rows  for  the Additive Group Testing
described above. However, it's not clear whether their probabilistic construction can be derandomized, and 
at which cost. We thought   worthwhile to mention  that our combinatorial tool gives, for free,
a solution to the Additive Group Testing problem using  number of tests that differ from
the optimal one for only  a factor of $\log p$. 
\smallskip

\noindent
{\bf Monotone Encodings.}
Moran {\em et al.}~posed the problem of efficiently constructing  $(n,k)$-monotone encodings of size $r,$ 
(denoted by $ME(n,k,r)$), i.e., 
monotone injective functions mapping subsets of $[n],$ of size up to $k,$  into $2^{[r]}$  \cite{Moran}.
Monotone encodings are relevant to the study of tamper-proof data 
structures and arise also in the design of broadcast schemes in certain communication networks
A simple counting argument shows that 
$ME(n,k,r)$ can only exist for $r = \Omega(k \log n/k).$ 
We can use our \newsel\ for obtaining $ME(n,k, O(k \log n/k))$ in the following way.
Let $M^{[t]}$ denote the $(t, \v, n)$-\newsel\ defined by the vector $\v$ whose $i$th component is $v_i = \lfloor i/ 2 \rfloor+1$ for each 
$i=1, \dots, t.$  By Lemma \ref{lemma:supsel_prop}, we have that for any $S \subseteq [t/2],$ from ${\bf a}_S$ we can identify 
at least $|S|/2$ of the columns in $M^{[t]}(S).$
Let $S^{yes}$ (resp. $S^{no}$ be the subset of these columns which  we can (resp. cannot) 
identify from ${\bf a}_S.$

We can obtain our mapping in the following way.
Given $S_0 \in {[n] \choose  {\leq k}},$ we map it to the concatenation of the vectors ${\bf a}_{0} {\bf a}_{1}, \dots, ..., {\bf a}_{\log k},$
where ${\bf a}_{i}$ is the Boolean sum of the columns of $M^{[k/2^{i-1}]}(S_i),$ with  $S_i = S_{i-1}^{no}.$  

The mapping is of size $\sum_{j=0}^{\log k} \frac{2k}{2^j} \log \frac{n 2^j}{2k} = O(k \log n/k)$,
therefore of optimal size. Moreover, by observing that for each $S \subseteq T$
we have ${\bf a}_{S} \leq {\bf a}_{T}$ and $S^{no} \subseteq T^{no}$, we also have that the mapping is monotone.
By our Theorem \ref{theo:main} such mapping can be deterministically computed in $O(n^k)$-time.

Alon and Hod \cite{Alon-Hod} defined $(k, \alpha)$-FUT families in order to obtain 
$ME(n,k, O(k \log \frac{n}{k}))$ in a way analogous to the one we depicted above, i.e, by chaining
 $(\frac{k}{2^t}, \frac{1}{2})$-FUT families\footnote{In fact, via  \newsel{\sc s}, 
 we can  provide 
 constructions of  optimal size 
 $(k, \alpha)$-FUT families, for any $1/2< \alpha < 1 - \frac{1}{k}.$}
  of cardinality $n$ for 
$t=0,1,\dots, \log k.$ However,  for  optimal, i.e., $O(k \log n/k)$-size monotone encodings
no explicit deterministic construction  has been provided so far  \cite{Alon-Hod,Moran}.

\smallskip

\noindent
{\bf Selector-based data compression.}  \label{sec:GTAPP}
Let $M$ be a $(p+1, 2p, n)$-selector of
size $m=O(p\log(n/p)).$ 
Let $\x$ be a binary vector with $||\x||_0\leq p.$ 
Define  the encoding of $\x$ as the vector $\y$ equal to the componentwise
OR of columns of $M$ corresponding
to the positions of the $1$'s in $\x$.
Let $x_{i_1}, \ldots, x_{i_d}$, $d\leq p$,
be all the components of $\x$ such that
$x_{i_1}= \ldots= x_{i_d}=1$.
By Lemma \ref{lemma:supsel_prop}, 
there exist at most
$t$ other columns $m_{j_1}, \ldots, m_{j_t}$ of matrix $M$,
$t\leq p$, such that
$\y=m_{j_1} \vee \ldots \vee m_{j_t} \vee
m_{i_1} \vee \ldots \vee m_{i_d}.$

Now, think of  an ``encoder'' that works
as follows: for a given vector $\x$ it first computes
its  encoding
$\y$, then it computes $A=\{i_1, \ldots , i_d\},
B=\{j_1, \ldots , j_t\}$, and subsequently  it computes
an ordered list $L$ from
$A\cup B$. Finally, the encoder computes
a binary vector $\z$ of
length $2p$ such that
$z_k=1$ if and only if the $k$-th element
of the ordered list $L$ is an element of $A$.
The encoding of $\x$ is now the concatenated binary
vector $\y\z$ of length $O(p\log(n/p))+2p=O(p\log(n/p))$.
One can see that $\x$ can be (efficiently) recovered
from $\y\z$ and that the length of the encoding
$\y\z$ of $\x$ is information
theoretically optimal.

An extension of the above reasoning can be
carried out also to a
scenario where $\x$ is generated by
a probabilistic source, provided
that $Pr\{||\x||_0>p\}$ goes to zero
as the length $n$ of $\x$ grows.

The above  encoding procedure has  some features which might be of some
 interest in the area of data compression. Specifically, it does not 
require  construction of  code dictionary, nor it is based on  statistical 
analysis of the sequences to be compressed. Moreover, the encoding/decoding  procedure only involves 
simple operations on Boolean vectors (OR's of them and checks for containments),
 which leads   to  fast implementation.
Furthermore, the above procedure provides a faster  alternative for optimal size  enumerative encoding 
of low-weight binary sequences.
\cite{Cov,Ryabko}. In particular, for binary vectors of Hamming weight at most $d,$ 
our encoding/decoding procedures require
time $O(nd\log (n/d)$), whereas the procedures given in \cite{Ryabko}
require time $O(n\log^2 n\log\log n)$ for the encoding, and
time $O(n\log^3 n \log\log n)$ for the decoding.

\smallskip

\noindent
{\bf Tracing many users (or finding many positives).}
In \cite{LR} the authors introduced $k$-out-of-$r$ Multi User Tracing families, aka $MUT_k(r).$ 
A family ${\cal F}$ of $n$ many subsets of $[m]$ is $MUT_k(r)$ if 
given the union of $\ell \leq p$ of the sets in ${\cal F},$ one is able to identify  at
least $k$ of them, or all if $\ell < k.$ Such definition is motivated by applications in multiple access channel communication and
DNA computing (see \cite{LR} and references quoted therein).

In \cite{Alon-Asodi} it was proved that $MUT_k(r)$ families exist  for $m=O((r+k^2)\log\frac{n}{r}),$ determining the maximum possible 
{\em rate} $\frac{\log n}{m}$ for all $k \leq \sqrt{r}$ up to a constant factor. Somehow surprisingly, 
in all this range the rate is $\Theta(\frac{1}{r}),$ independently of $k.$ 
However, no constructive proof of such ``optimal'' rate families has been provided so far.  

We can use our  
\newsel{\sc s}  to match such result:
Let  $M$ be a 
$(2r, \v, n)$-\newsel\, where the vector ${\bf v} = (v_1, \dots, v_{2r})$ is defined by:
$v_i = i$ for $i=1, \dots, k$; $v_i = k,$ for $i=k+1, \dots, 2r-1,$ and $v_{2r} = r+1.$
\remove{
  $$v_i = \begin{cases}
  i & if \;\; i \leq k \cr
  k & if \; \; k+1 \leq i \leq 2r-1 \cr
  r+1 & if \;\; i = 2r.
  \end{cases}
  $$
}

First, we notice that $M$ is a $(k,k,n$)-selector, i.e., a $(k-1)$-superimposed code, hence 
every union of up to $k-1$ columns is unique. Moreover, for any $k \leq \ell \leq r,$ by Lemma \ref{lemma:supsel_prop} we have that 
at least $k$ columns out of $\ell$ can be identified by their Boolean sum. 
These two properties show that the sets whose indicator vectors coincide with the columns of $M,$ 
form an  $MUT_k(r)$ family. 
Therefore, Theorem  \ref{theo:main} applied to $M$ provides
the best known bound on the size of $MUT_k(r)$ families, 
i.e., the $O(\max\{r, k^2\} \log n/r)$ of \cite{Alon-Asodi}. Our main theorem also explicitly shows that 
the result of  \cite{Alon-Asodi} can be attained by a constructive 
$O(n^k)$ strategy.

\smallskip

\noindent
{\bf The $(d, \ell)$-list disjunct matrices.}
Indyk {\em et al.} \cite{soda10} studied  $(d,\ell)$-list disjunct matrix which are $m \times n$ binary matrix 
such that the following holds:
for any disjoint subsets $S, T$ of columns, such that $|S| \leq d$ and
$|T| \geq \ell,$ there exists a row where there is a  $1$ among the columns in $T,$
while all the columns  in $S$ have a $0.$ Such structure
was also considered in \cite{TCS,DGV,Eppstein,Cheraghchi}.

One can easily verify that a $(d+\ell, d+1, n)$-selector is also a $(d,\ell)$-list disjunct matrix.  
As a consequence, our Lemma \ref{lemma:selectors} (below) provides  improved bounds on construction of  
$(d, \ell)$-list disjunct matrices\footnote{Analogous bounds, in terms of size, are derivable 
from  \cite{DGV} via $(p,k,n)$-selectors.   However, their construction time
is exponential in $n.$}
 compared to the ones given in  \cite{soda10}.  

For any $d \geq \ell,$ by using $(d + \ell, d+1, n)$-selector, we obtain 
  $(d, \ell)$-list disjunct matrices of size $O(\frac{(d+\ell)^2}{\ell} \log \frac{n}{\ell})$
for any constant $d$ and $\ell.$ This improves on  \cite{soda10}, particularly for $d$ large compared to $\ell.$
Also for $\ell = \Theta(d)$ and 
particularly for  $(d,d)$-list disjunct matrices our bound  compares favorably  
with the $O((d \log n)^{1+o(1)})$ size bound given in  \cite{soda10}
and the $O(d^{1+o(1)} \log n))$ size bound given in  \cite{Cheraghchi}. 
Alternatively, for  $d< \ell$  one can see that  a $(2d, d+1, n)$ selector is also a $(d, \ell)$-list disjunct matrix. 
Such a selector can be  constructed of size $O(d \log n/d),$ in time $n^{2d+o(1)}.$

We remark that the above results on the size of 
$(d, \ell)$-list disjuct matrices via 
selectors, are tight with respect to the lower bounds provided 
in \cite[Theorem 2]{DGV}, as reported in Table \ref {table:comparisons}.

\section{Bounds on the size of  a $(p, \v, n)$-\newsel} \label{sec:construction}  \label{sec:derandomization}

In this section we prove the bound on the size of a $(p, \v, n)$-\newsel\ as announced in Theorem \ref{theo:main}. 
First we present an immediate lower bound following from the ones 
 of \cite{Chlebus,DGV} on the size of $(p,k,n)$-selectors. 

\begin{theorem} \label{theo:lower_bound}
The size of a $(p, \v, n)$-\newsel\ has to be 
$$\Omega\left(\max_{j=1, \dots, p} \frac{j^2}{j-v_j+1} \frac{\log(n/j)}{\log\left(j/(j-v_j+1)\right) + O(1)} \right).$$
\end{theorem}
\remove{
\begin{proof}
By definition, a  $(p, \v, n)$-\newsel\ simultaneously a $(v_j, j, n)$-selector, for each $j=1, \dots, p.$ 
Therefore, obviously, the \newsel\'s size is at least as large as the size of the largest  
$(v_j, j, n)$-selector it includes, over all $j=1, \dots, p.$ The desired result now directly follows from 
\cite[Theorem 2]{Chlebus}, which states that any $(v_j, j, n)$-selector has size 
$\Omega\left(\frac{j^2}{j-v_j+1} \frac{\log(n/j)}{\log\left(j/(j-v_j+1)\right)+O(1)} \right).$
\end{proof}
}

For the upper bound, we first give a proof based on the probabilistic method and then derandomize it. 
We need the following two lemmas. 

\begin{lemma} \label{lemma:strong_selectors}
There exists a $(p,\v,n)$-\newsel\ of size
$$m = O\left(\max_{j=1, \dots, p} \frac{3pej}{(j-v_j + 1)} \log (n/j)\right).$$
\end{lemma}
\begin{proof}
Generate the $m\times n$ binary
matrix $M$ by choosing each entry randomly and independently,
with $Pr(M[i,j]=0)=(p-1)/p=x$.
Fix an integer $j \leq p.$ Fix $S \in {[n] \choose j}.$ For any subset
$R$ of $j-v_j+1$ rows of $I_j$ 
let $E_{R,S}$ be the event that the submatrix $M(S)$ does not 
contain \textit{any}   of the $(j-v_j+1)$ rows of $R$.
We have
\begin{equation}\label{eq:ERS}
Pr(E_{R,S})=\left(1-(j-v_j+1)x^{j-1}(1-x)\right)^m
\end{equation}
Let $R_1, \ldots , R_t$, $t={j\choose{j-v_j+1}}$
be all possible subsets of exactly $j-v_j+1$ rows 
of the matrix $I_j$, and let $N_S$ be the event that, for some index $i \in \{1, \dots, t\},$ 
the sub-matrix $M(S)$ does not contain \textit{any} of the 
rows of the subset $R_i$. By the union bound 
we have
\begin{eqnarray}\label{eq:ES}
Pr(N_{S})&=&Pr\left(\bigvee_{i=1}^t E_{R_i,S}\right) 
\leq  {j\choose{j-v_j+1}}  \label{eq:ES_2}
\left(1-(j-v_j+1)x^{j-1}(1-x)\right)^m
\end{eqnarray}
One can see that $N_S$ coincides with the 
the event that the sub-matrix
$M(S)$ 
contains
strictly less than  $v_j$ rows of  $I_j$.
To see this, it is enough to observe that   if $M(S)$ contains less than $v_j$ 
rows of $I_j$ it means that there is some $i$ such that 
$M(S)$ does not contain any of the rows in $R_i.$ 

Let $Y_M$ denote the event that the matrix $M$ is a $(p, \v, n)$-\newsel. 
We can use again the union bound 
to estimate the probability of the negated event $\overline{Y_M}.$
If $M$ is not a  $(p, \v, n)$-\newsel\ then 
there exists an integer $j \in [p]$ such that for some $S \in {[n] \choose j}$  the event $N_S$ happens. 
Therefore, 
$$Pr(\overline{Y_M}) = Pr\left(\bigvee_{j=1}^p \bigvee_{S \in {{[n]} \choose j}} N_S\right),$$ whence,  we obtain:
\begin{equation} \label{eq:strong_selector_prob}
Pr(Y_M) \geq 1 - \sum_{j=1}^p {n \choose j} {j \choose {j-v_j + 1}} \left(1 - \left( j-v_j + 1\right) x^{j-1} (1-x)\right)^m.
\end{equation}

By the probabilistic method, there exists a $(p, \v, n)$-\newsel\ of size $m^* = {\rm argmin}_{m \geq 1}~Pr(Y_M) > 0.$ 
The rest of the proof will consist in showing that $m^*$ satisfies the bound claimed.  



Let us focus on the value $c_j$ such that the $j$-th summand in (\ref{eq:strong_selector_prob}) satisfies the following inequality

\begin{equation} \label{eq:j_summand}
{n \choose j} {j \choose {j-v_j + 1}} \left(1 - \left( j-v_j + 1\right) x^{j-1} (1-x)\right)^{c_j  j \log n/j} \leq 1/p
\end{equation}


We shall use the following two  inequalities
\footnote{A step by step computation is in the appendix}
\begin{equation} \label{eq:asym1}
\left(1- \left( j-v_j +1 \right)x^{j-1} (1-x) \right)^{c_j j \log (n/j)} \leq 
\left(\frac{n}{j} \right)^{-\frac{ (j-v_j +1)c_j j}{ep}}
\end{equation}
\begin{equation} \label{eq:asym2}
{n \choose j} {j \choose {j-v_j + 1}} \leq 
 n^{j} 2^{\frac{j}{2}} e^{\frac{3j}{2}} j^{-j} 
\end{equation}

By  (\ref{eq:asym1})-(\ref{eq:asym2}), we have that the left-hand-side of (\ref{eq:j_summand})  can be upper bounded by 
\begin{equation} \label{eq:asym3}
 n^{j -\frac{c_j \left(j-v_j + 1\right) j}{ p e}}  
2^{\frac{j}{2}} e^{\frac{3j}{2}}
j^{-\left(j - \frac{c_j \left(j-v_j + 1\right) j}{ p e} \right)} = 
n^{j - {\frac{c_j \left(j-v_j + 1\right) j}{ p e}}} 2^{\frac{j}{2}} e^{\frac{3j}{2}} j^{-j + {\frac{c_j \left(j-v_j + 1\right) j}{ p e}}},
%
%
\end{equation}

Therefore, if we take   
$c_j  = \frac{3 p e}{(j-v_j + 1)}$ we have that (\ref{eq:asym3}) can be further upper bounded with
$ n^{-2j} e^{2j} j^{2j}$ which is not larger than  $1/p$
for all $n \geq 20$ and $n > p \geq j > 0.$
Therefore, by taking 
\begin{equation} \label{eq:m*}
m = \max_{j=1, \dots, p} c_j \log(n/j) =  \max_{j=1, \dots, p} \frac{3 p e j}{(j-v_j + 1)}  \log \frac{n}{j}
\end{equation}
we can have each of the summands in (\ref{eq:strong_selector_prob}) smaller than $1/p,$ 
hence guaranteeing  $Pr(Y_M)>0.$ By definition $m^* \leq m$ which  concludes the proof. \qed
\end{proof}

The same analysis as above, tailored for a $(p,k,n)$-selector gives the following bound, whose proof is 
deferred to the appendix.

\begin{lemma} \label{lemma:selectors}
For each $0 \leq k < p < n,$ there exists a $(p, k, n)$-selector of size 
\begin{equation}\label{UB_selector}
m = \left( \log_2 \frac{e}{e-1+\frac{k}{p}}\right)^{-1} p \log \frac{n}{p} \left(1+o(1) \right)
 \leq   \frac{2 p^2}{p-k+1} \log \frac{n}{p} \left(1+o(1) \right).
\end{equation}
Moreover, there exists a $(p, p, n)$-selector of size 
$\displaystyle{m = \frac{e p^2}{\log_2 e} \log (n/p)\left(1 + o(1)\right)}.$
\end{lemma}

We can now combine the last two lemmas to obtain the main result of this section, providing an almost tight upper bound on the 
size of a \newsel. 

\begin{theorem} \label{theo:strong_sel}
There exists a  $(p,\v,n)$-\newsel\ of size
%
%
$$m = O(\max_{j=1, \dots, p} k_j \log (n/j)), \qquad \mbox{where } 
k_j =  \min\left\{\frac{3pej}{(j-v_j + 1)}, \frac{e j^2}{\log_2 e} \right\}$$
\end{theorem}

\begin{proof}
Fix  
$k = \max \left\{ j \mid \frac{3 p e j}{(j-v_j + 1)} > \frac{e j^2}{\log_2 j} \right\}.$
 Let $M_1$ be a minimum size  $(k, k, n)$-selector.  In particular this is a 
$(k, <1,2,\dots, k>, n)$-\newsel\, hence {\em a fortiori} it is also a $(k, (v_1, \dots, v_k), n)$-\newsel.

Let $M_2$ be a minimum size $(p, (0,\dots, 0, v_{k+1}, \dots, v_p), n)$-\newsel. 

Let $M$ be the binary matrix obtained by pasting together, one on top of the other, $M_1$ and 
$M_2.$  It is not hard to see that  $M$ is a  $(p, \v, n)$-\newsel. By 
Lemmas \ref{lemma:selectors} and \ref{lemma:strong_selectors},  $M$ satisfies the desired bound. 
The proof is complete. \qed
\end{proof}

\begin{remark}  
Note that, if there exists a constant $\alpha$ such that $v_j \leq \alpha j$ for each $\sqrt{p} < j \leq p,$ then the size of the 
\newsel\ is $O(p \log \frac{n}{p}),$ 
matching the information theoretic lower bound.  Particular cases are given by instances where for each $j,$ we have 
$v_j = f_j(j)$ for some function $f_j$ such that  $f_j(j) = o(j).$
\end{remark}

\noindent
{\bf Deterministic construction.} By using  the method of the conditional expectations 
(see, e.g.,  \cite{Mitzenmacher}) we can  derandomize the result of the previous section and 
provide a deterministic  construction of the   $(p, \v, n)$-\newsel\ of Theorem \ref{theo:strong_sel} which is polynomial in $n$ but
exponential in the second parameter $p.$
More precisely we obtain the following result, whose proof  is deferred to 
the appendix.

\begin{theorem} \label{theorem:strong_construction}
There exists  a deterministic $O\left( p^3 n^{p+1} \log n  \right)$ construction 
of the $(p,\v,n)$-\newsel\ given by Theorem  \ref{theo:strong_sel}.
\end{theorem}

  \newpage
  \centerline{\bf \large APPENDIX}
  
  \appendix

\section{The proof of Theorem \ref{theo:lower_bound}}

\noindent
{\bf Theorem \ref{theo:lower_bound}.}
{\em The size of a $(p, \v, n)$-\newsel\ has to be 
$$\Omega\left(\max_{j=1, \dots, p} \frac{j^2}{j-v_j+1} \frac{\log(n/j)}{\log\left(j/(j-v_j+1)\right) + O(1)} \right).$$}
\begin{proof}
By definition, a  $(p, \v, n)$-\newsel\ simultaneously a $(v_j, j, n)$-selector, for each $j=1, \dots, p.$ 
Therefore, obviously, the \newsel\'s size is at least as large as the size of the largest  
$(v_j, j, n)$-selector it includes, over all $j=1, \dots, p.$ The desired result now directly follows from 
\cite[Theorem 2]{Chlebus}, which states that any $(v_j, j, n)$-selector has size 
$\Omega\left(\frac{j^2}{j-v_j+1} \frac{\log(n/j)}{\log\left(j/(j-v_j+1)\right)+O(1)} \right).$
\end{proof}

\section{The calculations for inequalities (\ref{eq:asym1}) (\ref{eq:asym2})}

As  regards (\ref{eq:asym1}), we have 
\begin{eqnarray*} 
\left(1- \left( j-v_j +1 \right)x^{j-1} (1-x) \right)^{c j \log (n/j)} &=& 
\left(1- \left( j-v_j +1 \right)\frac{(p-1)^{j-1}}{p^j} \right)^{c j \log (n/j)}\\
&=& \left(1- \frac{\left( j-v_j +1 \right)}{p}\left(1-\frac{1}{p}\right)^{j-1} \right)^{c j \log (n/j)}\\
&\leq& \left(1- \frac{\left( j-v_j +1 \right)}{p}\left(1-\frac{1}{p}\right)^{p-1} \right)^{c j \log (n/j)}\\
&\leq& \left(1- \frac{\left( j-v_j +1 \right)}{p}e^{-1} \right)^{c j \log (n/j)}\\
&\leq& e^{-\frac{ (j-v_j +1)c j \log (n/j) }{ep}}\\
&=& \left(\frac{n}{j} \right)^{-\frac{ (j-v_j +1)c j}{ep}}.
\end{eqnarray*}

As regards (\ref{eq:asym2}), we only need to use the known inequalities ${a \choose b} \leq (\frac{ae}{b})^b$ and 
${a \choose b} \leq {a \choose {a/2}},$ so we have

\begin{equation*} 
{n \choose j} {j \choose {j-v_j + 1}} \leq {n \choose j} {j \choose {j/2}} \leq
\left(\frac{n e}{j} \right)^{j} \left(2 e\right)^{j/2} = 
n^{j} 2^{\frac{j}{2}} e^{\frac{3j}{2}} j^{-j} 
\end{equation*}

\section{The  construction of the   $(p, \v, n)$-\newsel} 

In this appendix we prove Theorem \ref{theorem:strong_construction} presenting a derandomized construction of 
the $(p, \v, n)$-\newsel\ of Theorem \ref{theo:strong_sel}.

We shall need the following technical fact whose proof is  immediate. 

\begin{fact} \label{defi:partial_matrix_prob}
Fix integers $m, p \geq 1$ and $0 \leq k' \leq k \leq p.$   
Let $A$ be a subset of $k$ distinct rows of the identity matrix $I_p. $ 
Let $x = (p-1)/p$ and  $M$ be a randomly generated $m \times p$ binary matrix with 
each entry being independently chosen to be $0$ with probability $x.$
 Let $f(m,k', k)$ denote the probability that 
at least $k'$ distinct rows of $M$ are in  $A.$  Then, it holds that
%
%
%
%
%
 $$f(m, k', k) = 
 \begin{cases}
 (1- \alpha k) f(m-1, k', k) + \alpha k f(m-1, k'-1, k-1) & \mbox{if } m \geq k' > 0 \\
 1 
  & \mbox{if } k' = 0\\
 0 & \mbox{if } m < k',
\end{cases}
 $$
where $\alpha = x^{p-1}(1-p)$ is the probability of generating a particular row of $A.$ 

By using the above expression, we  can compute in $O(k^2 m)$ time 
 the complete table of values $f(a,b,c)$
 for each $a=1, \dots, m, \, b = 1, \dots, k', \, c=1, \dots, k.$ 
\end{fact}

We limit ourselves to discuss the derandomization of the proof of Lemma \ref{lemma:strong_selectors}. 
The same  ideas can be used to derandomize the construction of the selectors provided by Lemma \ref{lemma:selectors}, which are
needed to construct the \newsel\  satisfying Theorem \ref{theo:strong_sel}.

For each $j=1, \dots, p$ and each subset $S$ of $j$ columns of $M$ let $X^{(j)}_S$ be the indicator random variable which is 
equal to $1$ if $M(S)$ contains at
least $k$ rows of the identity matrix $I_j.$ 
 %
Let $X = \sum_{j=1}^p \sum_{S \in {[n] \choose j}} X^{(j)}_S.$  
It follows that  $E[X] = \sum_{j=1}^p \sum_{S \in {[n] \choose j}} Pr (X^{(j)}_S = 1).$
Since $Pr(X^{(j)}_S = 1) = Pr(\overline{N_S}),$ by  (\ref{eq:ES}), we have 
\begin{equation} \label{eq:E(X)}
E[X] 
\geq \sum_{j=1}^p {n \choose j}\left(1 - {j \choose {j-v_j+1}} \left(1-(j-v_j + 1) x^{j-1}(1-x)\right)^{m} \right)
\end{equation}
From Section \ref{sec:construction}, in particular from equation (\ref{eq:j_summand}), 
we know that  the  choice of $m$ satisfying (\ref{eq:m*}), guarantees 
$${j \choose {j-v_j + 1}}\left( 1-(j-v_j + 1) x^{j-1}(1-x) \right)^{m} < \frac{1}{p{n \choose j}}.$$
This, together with  (\ref{eq:E(X)}) gives  
\begin{equation} \label{eq:start_app}
E[X] > \sum_{j=1}^p {n \choose j} \left(1 - \frac{1}{p{n \choose j}} \right) = \sum_{j=1}^p {n \choose j} - 1.
\end{equation}

This quantity represents the expected total number of 
sub-matrices of $j$ columns (summed up over all $j=1, \dots, p$) with at least  
$v_j$ rows of the identity matrix $I_j,$ assuming each entry being $0$ with probability  $(p-1)/p.$
 
We now choose the entries of $M$ one at a time, trying to maximize the above expectation conditioned on  the 
entries already chosen. We shall construct $M$ row by row.  $M[r,c]$ will denote the entry in row $r$ and column $c.$
Once the entry $M[r,c]$ has been fixed, we use  
$\mu_{r c}$ to denote its  value.

For each $r=1, \dots, m,$ and $c = 1, \dots, n,$ let $\X[r,c]$ be the expected value of $X$ 
conditioned on the choices of the entries made 
before chosing entry $(r,c).$ Also, let $\X_0[r,c]$ and $\X_1[r,c]$ be the same but also conditioned to $M[r,c] = 0$ or $M[r,c] = 1$ respectively.
Let $\prec_{lex}$ denote the lexicographic order among pairs, i.e., $(x,y) \prec_{lex} (x',y')$ iff $x , x'$ or $x = x'$ and $y < y'.$

We have  
\begin{eqnarray*}
\X[r,c] &=& \sum_{j=1}^p \sum_{S \in {[n] \choose j}}  Pr\left(X^{(j)}_S = 1 \mid M[r',c'] = \mu_{r' c'}, \mbox{for each } (r', c') \prec_{lex} (r, c)\right).\\
\remove{
This is the expectated value of $X$ before the entry $M[r,c]$ is chosen, i.e., under the following two conditions:
\begin{itemize}
\item  
$M[i', j'] = \mu_{i', j'},$ for each  $(i', j') \prec_{lex} (i, j)];$ (these are the values already fixed)
\item the remaining entries are $0$ with probability $\frac{p-1}{p}.$
\end{itemize}

Also, it is comfortable to define the same expectation conditioned also on the two possible choices of $M[r,c],$ i.e.,
}
\X_0[r,c] &=& \sum_{j=1}^p \sum_{S \in {[n] \choose j}}  Pr\left(X^{(j)}_S = 1 \mid M[r',c'] = \mu_{r' c'}, \mbox{for each } (r', c') \prec_{lex} (r, c) 
 \mbox{ and } M[r,c] = 0\right)\\
\X_1[r,c] &=& \sum_{j=1}^p \sum_{S \in {[n] \choose j}}  Pr\left(X^{(j)}_S = 1 \mid M[r',c'] = \mu_{r' c'}, \mbox{for each } (r', c') \prec_{lex} (r, c) 
 \mbox{ and } M[r,c] = 1\right).
 \end{eqnarray*}

In accordance to the method of the conditional expectations we set  $M[r,c] = 0$ if and only if $\X_0[r,c] \geq \X_1[r,c].$
 
 
It is not hard to see that  this leads to the construction of  the desired selector. 
\remove{
Suppose that we are about to  choose the value $M[i,j].$
Under the  hypothesis that 
 the remaining values are chosen to be $0$ with probability $\frac{p-1}{p},$ we have 
 
 $$\X[i,j] = \frac{p-1}{p} \X_0[i,j] + \frac{1}{p} \X_1[i,j].$$
 On the left hand side of the equation, we have the expectation (on the number of $M(S)$ with $k$ rows from $I_p$) as computed immediately before choosing $M[i,j]$ and after we have chosen $M[i, j-1].$ The choice of $M[i,j-1]$ is done by maximizing between  
$\X_0[i,j-1] $ and  $\X_1[i,j-1].$ 

Therefore, we have 
 }
 We have 
 $$\max\{\X_0[r,c-1] , \X_1[r,c-1] \} = \X[r,c] = \frac{p-1}{p} \X_0[r,c] + \frac{1}{p} \X_1[r,c] \leq \max \{ \X_0[r,c], \X_1[r,c]\} = \X[r,c+1],$$
 where the first and the last equality follows by the definition of the strategy and  the second equality by the definition of conditional expectation. 
 
This shows that   the expectation $\X[r,c]$ is monotonically increasing.\footnote{For the sake of the presentation, 
 we are here tacitly assuming that 
 $1<c<n.$ It is not difficult to extend the argument also for the extreme cases when $c \in \{1, n\},$ i.e., 
 the expectations involved are about consecutive rows of $M.$.} 
 By (\ref{eq:start_app}), we start with $\X[1,1] > \sum_{j=1}^p {n \choose j} - 1.$ 
 Moreover, once all the entries have been chosen, the above expectation is in fact the actual number of submatrices satisfying the 
 \newsel\ conditions. This must be 
 an integer and by the starting condition and the above monotonicity it is greater than $\sum_{j=1}^p {n \choose j} - 1,$ 
 which means that the  matrix 
$M$ we have so constructed is indeed a $(p,k,n)$-selector.
 
 
 We also have to show that we can compute $\X_0[r,c]$ and $\X_1[r,c]$ ``efficiently''. Let us fix $j$ and focus on a single subset  
 $S$ of $j$ columns  and the corresponding sub-matrix $M(S).$ It will be enough to show that we can 
 compute efficiently  the following quantity:
 $$\X^S[r,c] = Pr\left(X^{(j)}_S = 1 \mid M[r',c'] = \mu_{r' c'}, \mbox{for each } (r', c') \prec_{lex} (r, c)\right)$$
which is the probability of having at least $v_j$ rows of the identity matrix $I_j$ in the sub-matrix 
$M(S)$ given the choice made so far in $M$ up to 
$M[r, c-1].$ In fact, the computation of $\X[r,c], \X_0[r,c], \X_1[r,c]$ involves at most $\sum_{j=1}^p {n \choose j}$ probabilities $\X^S[r', c'].$

Suppose we are about to choose the value of $M[r,c].$ Let $a=a^S_{r-1}$ be the number of rows from $I_j$ which already appear in 
the first $r-1$ rows of $M(S),$ given the entries fixed so far.

\begin{enumerate}
\item If the $c$-th column of $M$ coincides with the $1$-st column of $M(S)$ then, no entry has been chosen 
so far in the $r$-th row of $M(S)$ and, recalling Fact \ref{defi:partial_matrix_prob}, it should not be difficult to see that we have 
$$\X^S[r,c] = f(m-r+1,v_j-a, j - a).$$  
 \item otherwise we have one of the following three cases
\begin{itemize}
\item[(i)] the $r$-th row of $M(S)$ cannot be one of the rows of $I_j$ which are not already in the first $r-1$ rows of $M(S)$, 
or there are already two entries with value $1.$ Therefore, 
$$\X^S[r,c] = f(m-r, v_j-a, j-a)$$
\item[(ii)] among the $c-1$ entries which have already been fixed in the $r$-th row of $M(S),$ 
there exists  exactly one entry which is equal to $1.$ Moreover,  there is exactly one choice of 
the remaining entries on row $r$ such that this row becomes one of the  $j-a$ rows of $I_j$ 
which do not appear among the  first $r-1$ rows of $M(S).$  In particular, if all the remaining entries of row $r$ are chosen to be $0$ then 
this becomes one of the row of $I_j$ not yet in $M(S).$ Therefore the probability that $M(S)$ ends up containing $v_j$ rows of $I_j$ 
becomes the probability that in the remaining $m-r$ there are at least $v_j-a-1$ rows from the $j-a-1$ not appearing
 in the first $r$ rows of $M(S).$Thus, 
$$\X^S[r,c] = x^{j-c+1}f(m-r, v_j-a-1, j-a-1) + (1- x^{j-c+1})f(m-r, v_j-a, j-a)$$
\item[(iii)]  there is no $1$ entry among the first $c-1$ entries already fixed in  row $r.$ 
Furthermore, among the $j-a$ rows of $I_j$ which are not in the  first $r-1$ rows of $M(S),$ 
there are exactly $b$ rows  which have only zeroes in the first $c-1$ positions. 
These are exactly the only rows  of $I_j$ which could appear in row $r$ of $M(S)$ 
given the choices made so far. If the $r$-th row of $M(S)$ ends up being one of these 
rows---which happens with probability $b x^{j-c}(1-x)$---then the probability of $M(S)$ containing 
$v_j$ rows from $I_j$ is the same as the probability of having $v_j-a-1$ rows out of the $j-a-1$ 
many which are not in the first $r$ rows of $M(S)$ in a random generated matrix with $m-r$ rows. 
Otherwise, the probability of having $v_j$ rows of $I_j$ in $M(S)$ is the same as the probability of having, 
in a randomly generate matrix of size $m-r,$ at least $v_j-a$ rows out of the 
$j-a$ which are not in the first $r-1$ rows so far chosen for  $M(S).$  Therefore, 
$$\X^S[i,j] = x^{j-c}(1-x) b f(m-r, v_j-a-1, j-a-1) + (1-b x^{j-c}(1-x))f(m-r, v_j-a, j-a)$$
 \end{itemize}
 
 \end{enumerate}

\subsection{Estimating  the time complexity of the derandomized strategy} \label{sec:Time_complexity}
 
 For each $r=1, \dots, m$ and $c = 1, \dots, n,$ the computation of the entry  $M[r,c]$ 
 requires that  for at most all the 
 $\sum_{j=1}^p {n \choose j}=O(p n^{p}$) sets  of columns, $S,$ we  
 look up a constant number of values of  $f(\cdot, \cdot, \cdot).$ 
 Recall that all such values  have been precomputed in time $O(p^2m).$ 
 Moreover, we need to keep track, for each $M(S),$ of the number $a^S_{r-1}$ 
 of rows of $I_j$ which already appear in the first $r-1$ rows of $M(S)$ 
 and   the $b$  rows of $I_j$ which coincide in the first $c-1$ bits with the first $c-1$ bits 
 of row $r$ of $M(S).$ By indexing, this can be easily done in $O(p).$ In total, we spend 
 $$O\left(n m \times p n^{p} \times p + p^2 \times m \right) = O\left( p^3 n^{p+1} \log n  \right).$$ 

This completes the proof of Theorem \ref{theorem:strong_construction}.


\section{The Proof of Lemma \ref{lemma:selectors}}

\subsection{Some useful estimates}
We shall need the following technical facts.

\begin{lemma} \label{lemma:const}
Fix an integer   $p > 1$ and let  $x = \frac{p-1}{p}.$ 

(a) For  $0<\epsilon < 1,$ it holds that 
$$\left(\log_2\frac{1}{\left(1 - \left( \epsilon p + 1\right) x^{p-1} (1-x)\right)}\right)^{-1} \leq \left(\log_2 \frac{e}{e-\epsilon} \right)^{-1}
< \frac{2 p}{1+p\epsilon}.$$
(b) Moreover (for $\epsilon = 0$) we have 
$$\left(\log_2\frac{1}{\left(1 -  x^{p-1} (1-x)\right)}\right)^{-1} \leq \frac{e\, p}{\log_2 e}.$$

\end{lemma}

\begin{corollary} \label{corollary:const}
Fix an integer   $p > 1.$ Then, for $x = \frac{p-1}{p},$ it holds that 
$$\left(\log_2\frac{1}{\left(1 - \left( \frac{p}{2} + 1\right) x^{p-1} (1-x)\right)}\right)^{-1} \leq 3.411.$$
\end{corollary}

\subsection{$(p,k,n)$-selectors exist of size $O(p \log \frac{n}{p})$: yet another proof!} \label{sec:prob_method}

Let $m,n, p\geq 1$ be integers
and  $M$ be an $m\times n$ binary matrix.
Recall that for each  $S\subseteq \{1,\ldots n\}, |S|=p$, we denote by
$M(S)$ the $m\times p$ submatrix of $M$ consisting of all
coloumns of $M$ whose indices are in $S$.


Fix integers $m,n, p\geq 1$ and generate a $m\times n$ binary
matrix $M$ by choosing each entry randomly and independently,
with $Pr(M[i,j]=0)=(p-1)/p=x$.
For any integer $k$, $1\leq k\leq p$, and for any subset
$R$ of $p-k+1$ rows of $I_p$ 
let $E_{R,S}$ be the event that matrix $M(S)$ does not 
contain \textit{any}   of the $(p-k+1)$ rows of $R$.
We have
\begin{equation}\label{eq:ERS_sel}
Pr(E_{R,S})=\left(1-(p-k+1)x^{p-1}(1-x)\right)^m
\end{equation}
Let $R_1, \ldots , R_t$, $t={p\choose{p-k+1}}$
be all possible subsets of exactly $p-k+1$ rows 
of matrix $I_p$, and let $E_S$ be the event that the 
sub-matrix $M(S)$ does not contain \textit{any} 
rows of some subset $R_i$. By the union bound 
we have
\begin{eqnarray}\label{eq:ES_sel}
Pr(E_{S})&=&Pr(E_{R_1,S}\vee \ldots \vee E_{R_t,S})\\
&\leq& {p\choose{p-k+1}}  \label{eq:ES_2_sel}
\left(1-(p-k+1)x^{p-1}(1-x)\right)^m =q
\end{eqnarray}
Let us denote by $N_S$ the event that the sub-matrix
$M(S)$ does not contain \textit{at least} $k$ rows of  $I_p$.
One can see that $Pr(N_S)=Pr(E_S)$. To see this, it is enough to observe that   
if $M(S)$ does not contain at least $k$ rows of $I_p$ it means that there is some $R_i$ such that 
$M(S)$ does not contain any of the rows in $R_i.$ 
Consequently
\begin{equation}\label{eq:q_sel}
Pr(N_S)\leq q
\end{equation}


There are ${n\choose p}$ events $N_S$, one for
each $S\subseteq \{1, \ldots ,n\}$ of cardinality 
$p$. 

Let $Y_M$ denote the event that the matrix $M$ is a $(p,k,n)$-selector. If $M$ is  $(p,k,n)$-selector it means that 
there exists no set $S$ such that  the event $N_S$ happens. We can use again the union bound 
to estimate the probability of the negated event $\overline{Y_M},$ as
$$Pr(\overline{Y_M}) = Pr\left(\bigvee_{S \in {{[n]} \choose p}} N_S\right),$$ whence,  we obtain:
\begin{equation} \label{eq:bound}
Pr(Y_M) \geq 1 - {n \choose p} {p \choose {p-k + 1}} \left(1 - \left( p-k + 1\right) x^{p-1} (1-x)\right)^m.
\end{equation}

Let $$m^* = {\rm argmin}_{m \geq 1}~
  Pr(Y_M) > 0.$$

One can conclude that there exists a  $(p,k,n)$-selector of size $m^*.$

We have to show that $m^* = O(p \log n/p).$ 

We can use  ${a \choose b} \leq (\frac{a e}{b})^b$ to bound the two binomial coefficients. We have
$$Pr(Y_M) \geq 1 - \left(\frac{n e}{p}\right)^p\left(\frac{p e}{p-k + 1}\right)^{p-k + 1} 
\left(1 - \left( p-k + 1\right) x^{p-1} (1-x)\right)^m.$$

The last quantity is positive for any $m$ such that
$$1 - \left(\frac{n e}{p}\right)^p\left(\frac{p e}{p-k + 1}\right)^{p-k + 1} 
\left(1 - \left( p-k + 1\right) x^{p-1} (1-x)\right)^m > 0,$$
which means 
$$\left(1 - \left( p-k + 1\right) x^{p-1} (1-x)\right)^m < 
\left(\left(\frac{n e}{p}\right)^p\left(\frac{p e}{p-k + 1}\right)^{p-k + 1} \right)^{-1},$$
i.e., 
$$m > \frac{\log_2\left(\left(\frac{n e}{p}\right)^p\left(\frac{p e}{p-k + 1}\right)^{p-k + 1} \right)}
{\log_2\frac{1}{\left(1 - \left( p-k + 1\right) x^{p-1} (1-x)\right)}} = 
\frac{p\log_2 \frac{n}{p} + (2p-k+1)\log e  + (p-k+1)\log_2 \frac{p}{p-k+1}}
{\log_2\frac{1}{\left(1 - \left( p-k + 1\right) x^{p-1} (1-x)\right)}}.$$

By Lemma \ref{lemma:const} we have that for any fixed  $0 < \alpha < 1$ and $k = \alpha p,$ we can bound $m^*$ as 
 $$m^* \leq \frac{1}{\log_2 e - \log_2(e - 1+ \alpha)} \left(p \log n/p + A_{p,k}\right),$$
 where $A_{p,k}$ is a constant only depending on $p$ and $k.$ I.e., we can find a $(p, k, n)$-selector of 
 size $O(p \log \frac{n}{p}).$ More precisely and  in the spirits of the lower bounds of \cite{Chlebus},  the estimates in  
 Lemma \ref{lemma:const} show that the size $m^*$ of the $(p,k,n)$-selector whose existence is guaranteed by the probabilistic method, is 
  bounded by 
 $$\frac{2 p^2}{p-k+1} \log \frac{n}{p}  \, (1+o(1)).$$ 
 
Notice that for $\alpha = 1,$ using Lemma \ref{lemma:const} (b) (with $\epsilon =  1-\alpha$),  we get the well known quadratic bound 
on the size of superimposed codes.

In some applications, as in the case of the $(d,d)$-list disjunct matrices,  of particular interest is
 the case $\alpha = \frac{1}{2}.$ For such case, using  Corollary \ref{corollary:const} we have
 $$m^* \leq 3.411 \left( p \log \frac{n}{p} +A_{p,k}\right),$$ which proves the desired result on the existence of  
a $(p/2,p,n)$-selector of size $3.411 p \log \frac{n}{p}(1+o(1)).$


\vspace{-0.5cm}

{\small 
\begin{thebibliography}{99}
\bibitem{Alon-Asodi} N. Alon and V. Asodi, 
{\em Tracing many users with almost no rate penalty}, 
IEEE Trans. on Information Theory, vol. 53, no. 1, pp. 437-439, 2007.
\bibitem{Alon-Hod} N. Alon, R. Hod, {\em Optimal Monotone Encodings}, 
IEEE Trans. on Information Theory, vol 55, no. 3, pp. 1343-1353, 2009.
\bibitem{balding}
 D.J. Balding  \textit{et al.} \textit{A
comparative survey of non-adaptive pooling design}, in Genetic
mapping and DNA sequencing,  IMA Volumes in Mathematics and
its Appl., T.P. Speed  \& M.S. Waterman (Eds.),
Springer-Verlag, 133--154, 1996.
\bibitem{char} { S. Chaudhuri and
 J. Radhakrishnan}, \textit{Deterministic restrictions in
circuit complexity}, in  Proc. of $28$th STOC, pp. 30--36,
   1996.
\bibitem{JCB08} Y. Cheng and D.Z. Du, {\em New Constructions of One- and Two-Stage Pooling Designs}, 
Journal of Computational Biology, 15 (2), pp. 195--205, 2008.
\bibitem{Cheng} Y. Cheng, D.Z. Du, G. Lin, 
{\em On the upper bounds of the minimum number of rows of disjunct matrices}, 
Optimization Letters, 3, pp. 297--302, 2009.
\bibitem{Chlebus} B.S. Chlebus and D.R. Kowalski, {\em Almost Optimal Explicit Selectors}, 
in Proc. of FCT 2005, LNCS 3623, pp. 270--280, 2005. 
\bibitem{Cheraghchi} M. Cheraghchi, {\em Noise-resilient group testing: Limitations and
constructions}, in Proc. of FCT 2009, 2009. 
\bibitem{CGR}M. Chrobak, L. Gasieniec, W. Rytter, \textit{Fast Broadcasting and Gossiping in Radio Networks}. FOCS 2000: 575-581
\bibitem{cms} { A.E.F. Clementi, A. Monti and R. Silvestri},
\textit{Selective families, superimposed codes, and broadcasting on
unknown radio networks}, in Proc. of Symp. on
Discrete Algorithms (SODA'01), 709--718, 2001
\bibitem{CM}G. Cormode and S. Muthukrishnan, \textit{Combinatorial Algorithms for Compressed Sensing},
 Proc. SIROCCO 2006, Lec. Notes in Comp. Sci.,   vol. 4056, pp. 280-294.
 \bibitem{Cov} T. Cover,
\emph{Enumerative source encoding}, 
IEEE Trans. Inf. Th.,
19, pp. 73--77, 1973.
\bibitem{Dam}
P. Damaschke, \emph{Adaptive versus Nonadaptive Attribute-Efficient Learning}. STOC 1998: 590-596.
\bibitem{TCS} A. De Bonis and U. Vaccaro, {\em Constructions of generalized 
superimposed codes with applications to group testing and conflict resolution in 
multiple access channels}, Theoretical Computer Science, {\bf 306}, pp. 223-243, 2003.
\bibitem{DGV}A. De Bonis, L. Gasieniec, and U. Vaccaro,
Optimal Two-Stage Algorithms for Group Testing Problems,
SIAM J. on Comp., vol. 34, No. 5, pp. 1253-1270, 2005.
\bibitem{du} { D.Z. Du and F.K. Hwang}, {\em Pooling Design and Nonadaptive Group Testing}, World Scientific, 2006.
\bibitem{DR} A.G. D'yachkov, V.V. Rykov, {\em Bounds of the length of disjunct codes},  
Problems Control Inform.  Theory 11, pp. 7--13, 1982. 
\bibitem{Dyachkov} A.G. D'yachkov,  V.V. Rykov, A.M. Rashad, {\em 
Superimposed distance codes},  Problems Control Inform. 
Theory,  18, pp. 237--250,  1989.
\bibitem{Eppstein} D. Eppstein, M.T. Goodrich, D.S. Hirschberg, {\em Improved Combinatorial 
Group Testing Algorithms for Real-World Problem Sizes}, SIAM J. on Comp., 36, pp. 1360-1375, 2007.
\bibitem{eff} { P. Erd\"os, P. Frankl, and Z. F\"uredi},
\textit{Families of finite sets in which no  set is covered by the union
of $r$ others}, Israel J. of Math.,   {\bf 51},
 75--89, 1985.
\bibitem{Ganguly} S. Ganguly, {\em Data stream algorithms via expander graph}, 
in Proc. of ISAAC 2008, pp. 52-63, 2008.
\bibitem{Gilbert} A.C. Gilbert, M.A. Iwen, M.J. Strauss, {\em Group Testing and Sparse Signal Recovery}, 
 $42$nd Asilomar Conf. on Signals, Systems, and Computers, pp. 1059-1063, 2008.
\bibitem{GK}V. Grebinsky and G. Kucherov,
{\em Optimal Reconstruction of Graphs under the Additive Model},
Algorithmica, vol. 28, no. 1, pp. 104-124, 2000.
\bibitem{ind} { P. Indyk}, \textit{Deterministic superimposed coding with
application to pattern matching},  Proc.  of $39$th FOCS,
 127--136, 1997.
\bibitem{soda10} P. Indyk, H.Q. Ngo, A. Rudra, 
Efficiently Decodable Non-adaptive Group Testing, in Proc. of $20$th SODA, pp. 1126-1142, 2010. 
\bibitem{kts} { W.H. Kautz and R.R. Singleton}, \textit{Nonrandom
binary superimposed codes}, IEEE Trans. on Inform. Theory,
{\bf 10}, 363--377, 1964.
\bibitem{black} { R. Kumar, S. Rajagopalan, A. Sahai},
  C\textit{oding constructions for blacklisting problems without computational assumptions},
in  Proc.\  of CRYPTO'99, 
609-623, 1999.
\bibitem{LR} B. Laczay and M. Ruszink\'o, 
{\em Multiple User Tracing Codes}, in Proc. of ISIT 2006, pp. 1900-1904, 2006.
\bibitem{linial} { N. Linial}, \textit{Locality in distributed graph
algorithms},  SIAM J. on Computing, {\bf 21},  193--201,
1992.
 Discrete Mathematics, vol. 162, pp. 311-312, 1996.
\remove{
\bibitem{Macula99a} A.J. Macula, 
{\em Probabilistic nonadaptive group testing in the presence of errors and DNA library screening},
 Annals of  Combinatorics, vol.  3, pp. 61-69, 1999. 
}
\bibitem{Mitzenmacher} M. Mitzenmacher and  E. Upfal, 
Probability and Computing: Randomized Algorithms and Probabilistic Analysis, 
Cambridge University Press, 2005.
\bibitem{Moran} T. Moran, M. Naor, G. Segev, {\em Deterministic history-independent strategies for storing
information on write-once memories}, in Proc. 34th ICALP, 303-315, 2007. 
\bibitem{Porat_FOCS} B. Porat and E. Porat,
{\em Exact and Approximate Pattern Matching in the Streaming Model},
in Proc. 50th FOCS, 315-323, 2009.
\bibitem{PR}E. Porat, and Rotschild, {\em Explicit non-adaptive combinatorial group testing schemes}, in Proc. of 
ICALP 2008, pp. 748-759, 2008.
\bibitem{Porat_ESA}  R. Clifford,  K. Efremenko, E. Porat, A. Rothschild,
{\em $k$-Mismatch with Don't Cares},
in Proc. 15th ESA,  151-162, 2007. 
%
\bibitem{Ryabko} B. Ryabko, {\em Fast Enumerative Source Conding}, in Proc. of 1995  IEEE Intern. Symp. on Inf. Th.\, p. 395.
\bibitem{Wolf} { J. Wolf}, \textit{Born again group testing: Multiaccess Communications},
IEEE Trans. Information Theory, {\bf 31}, 185-191, 1985.
\end{thebibliography}
}
\end{document}